\documentclass[aps,pra,floatfix,amsmath,amssymb,superscriptaddress]{revtex4}

\usepackage{latexsym}
\usepackage{float}
\usepackage{epsfig}
\usepackage{dcolumn}
\usepackage{amsmath}
\usepackage{amssymb}
\usepackage{graphicx}
\usepackage{times}
\usepackage{subfigure}
\usepackage[section]{placeins}
\usepackage{mathtools}
\usepackage{color}

\usepackage[T1]{fontenc}

\usepackage{grffile}
\usepackage[percent]{overpic}
\usepackage{microtype}
\usepackage[usenames,dvipsnames]{xcolor}
\usepackage[colorlinks,linkcolor=blue,citecolor=blue,urlcolor=blue]{hyperref}

\def\equationautorefname~#1\null{Eq.~(#1)\null}

\def\a{\alpha}

\def\g{\gamma}

\def\n{\nu}
\def\be{\begin{equation}}
\def\ee{\end{equation}}
\def\ba{\begin{eqnarray}}
\def\ea{\end{eqnarray}}
\def\la{\langle}
\def\ra{\rangle}
\def\a{\alpha}

\begin{document}

	\title{Investigation of the Behavior of Quantum Coherence in Quantum Phase Transitions of Two-Dimensional XY and Ising Models}
	\vspace{2cm}
	\author{N. Taghadomi}
	\affiliation{Department of Physics, Tarbiat Modares University, Tehran, Iran}
	
	\author{A. Mani}
	\affiliation{Department of Engineering Science, College of Engineering, University of Tehran, Iran}
	
	\author{A. Bakouei}
	\affiliation{Department of Physics, Tarbiat Modares University, Tehran, Iran}
	
	\vspace{2cm}

\begin{abstract}
	
We investigate the behavior of quantum coherence of the ground states of 2D Heisenberg XY model and 2D Ising model with transverse field on square lattices, by using the method of  Quantum Renormalization Group (QRG). We show that the non-analytic behavior of quantum coherence near the critical point, can detect quantum phase transition (QPT) of these models. We also use the scaling behavior of maximum derivative of quantum coherence, with system size, to find the critical exponent of coherence for both models and also the length exponent of the Ising model. The results are in close agreement with the ones obtained from entanglement analysis, that is while quantum coherence needs less computational calculations in comparison to entanglement approaches.

\end{abstract}

	
	\date{\today}
	
	\maketitle
	\vspace{0cm}
	
\section{Introduction}
	
Quantum coherence, as one of the most important properties of quantum systems, has attracted much attention in the last decade. That is due to the fact that quantum coherence is a key feature in many informational and computational tasks \cite{InfComp} and in different areas of physics such as spin models \cite{SpinModel1, SpinModel2, SpinModel3}, solid state physics \cite{SSPhys1, SSPhys2}, condensed matter \cite{Condensed matter}, quantum optics \cite{Quantum Optics}, quantum metrology  \cite{metrology}, quantum biology \cite{biology} and many other relevant fields. Due to the advantages that are based on this property, quantum coherence is regarded as a quantum resource and the resource theory of coherence is well established \cite{Quantum Resource Theories1,Quantum Resource Theories2}. Conditioned on the resource theory, various measures are defined to quantify quantum coherence \cite{Fidelity_Quantifying Coherence,Intrinsic Randomness,Measuring Concurrence,Measuring with Entanglement,Observable Measure,Quantum Fisher,basis independent measure}. For instance, the $l_{1}$ norm of coherence \cite{Quantifying Coherence} and the relative entropy of coherence \cite{Quantifying Coherence} are the earliest defined coherence measures.\\ 

On the other side, quantum phase transition (QPT) is a quantum phenomena which occurs at absolute zero temperature (or near that) and is due to quantum fluctuations of the ground state of system. In contrary to regular phase transitions which appear on temperature line, QPTs occur by varying some specific parameters of the Hamiltonian, like magnetic field or coupling constants \cite{Phase Transitions}.
For any QPT, there is a quantum critical point (QCP), where an abrupt change happens in the ground state, and thence some physical quantities are seen to have singular behaviors. Any measurable quantity which shows singular behavior in the QCP, can be regarded as a QPT-detector, and some critical exponents of the system can be derived by studying its changes near the critical point. \\
	
Recently a great interest is devoted to study the connections between QPTs and quantum mechanical features like entanglement \cite{entanglement and QPT,multi entanglement xxz model,Multipartite Entanglement and QPT,QPT nature}, quantum discord \cite{Q. Correlation} and quantum coherence \cite{QC_Sci,Dynamical Behavior Ising,Dynamics of XY,QPT Ising,QRG Khan,Q. C. spectrum,Various Q. M., Triangular1}. The sudden change of correlations near the QCP, leads to a singular behavior for some of the mentioned features, and hence they can be used as QPT-detectors. Specifically, quantum discord is shown to be an effective QPT-detector at finite temperatures, where entanglement is unable to show QPT because of thermal fluctuations \cite{Q. Correlation,Q. C. spectrum}. A preferable QPT-detector is the one which is easy to calculate or measure, and is able to detect the QPT of more systems, and also is useful to calculate critical exponents.
It has been shown  that quantum coherence is an appropriate quantity to study the QPT of some one dimensional models, such as Heisenberg XY, Ising, and XXZ Heisenberg spin-$\frac{1}{2}$ chains \cite{Dynamics of XY,QPT Ising,Dynamical Behavior Ising,QRG Khan}. That is while the only two dimensional model (up to our knowledge) which studies coherence in this context is \cite{Triangular1}, which considers QPT of two dimensional Ising model on a triangular lattice.\\

In fact most analysis of the properties of quantum systems in QPTs of higher dimensional systems are made by using Monte Calrlo simulations or other numerical methods \cite{Monte1, Monte2, Monte3, Monte4, Monte5, Monte6}, and exact diagonalization is only performed for some limited models like spin-1/2 ladder with four spin ring exchange \cite{exactdiag}. In contrary, the quantum renormalization group (QRG) method is an analytical technique for studying the ground state behaviors of large-site systems. The QRG method was used to exactly solve many of the above mentioned one dimensional models, for example the Ising model \cite{QPT Ising,Dynamical Behavior Ising} , XY \cite{Dynamics of XY}, XXZ \cite{Various Q. M.,QRG Khan,QRG Jafari,QRG XXZ}, and XYZ model in the presence of magnetic field \cite{QRGG2}, are studied completely in this context. The idea of QRG has also been applied to some limited  two dimensional systems, like Ising \cite{Triangular1, Q. E. 2D} and  XY models \cite{QRG XY, Two-Dimensional XY Model}, by dividing the lattice of spins into blocks which span the entire lattice. The authors of \cite{QRG XY, Two-Dimensional XY Model} have also used their QRG results to investigate the behavior of quantum entanglement near the critical point. \\

In this article, we use the approach of quantum renormalization group (QRG) to investigate if the $l_1$ norm of quantum coherence can be regarded as a preferable detector for QPT of two dimensional XY and Ising (with transverse field) Hamiltonians on square lattices. We show that, not only the coherence of multi-site states, but also the coherence of subsystems, even two site state, have non-analytic behavior in the vicinity of the critical point for high RG steps. Thence these coherences can show the QPT of the mentioned models.
We also demonstrate that the derivative of quantum coherence has a local maximum, which its position tends to the critical point for high RG steps.
The critical exponents and the correlation length are also derived by studying the scaling behavior of maximum derivative of $l_1$-norm of coherence near the QCP, and our results are in proper agreement with the ones obtained from entanglement studies and also exact values. That is while the calculations of quantum coherence is much easier than that of quantum entanglement. \\


The paper is organized as follows: In the next section, we briefly review the quantum renormalization method. The connections between quantum coherence and QPT are investigated in \autoref{XYMODEL} for the XY model in 2D, and in \autoref{Ising} for the Ising model with transverse field. Finally the paper ends with a conclusion in \autoref{conclusion}.\\

\section{Quantum Renormalization Group}\label{QRG}

It is not usually convenient to find the critical points and critical exponents by analytical eigenvalue solutions, and calculation by numerical methods is also computationally expensive.
Quantum renormalization method is an alternative analytical method which is generally used to study the quantum phase transitions and scaling behavior of many body systems \cite{QRGG2,QRG XXZ,QRG Jafari}. The idea is to use an iterative method for keeping the most important degrees of freedom of the system, while excluding the others. Here, by presenting the Kadanof's approach, we briefly review the general points of this method. \\

\begin{figure} [H]
	\resizebox{\linewidth}{!}{
		\begin{overpic}[height=8cm]{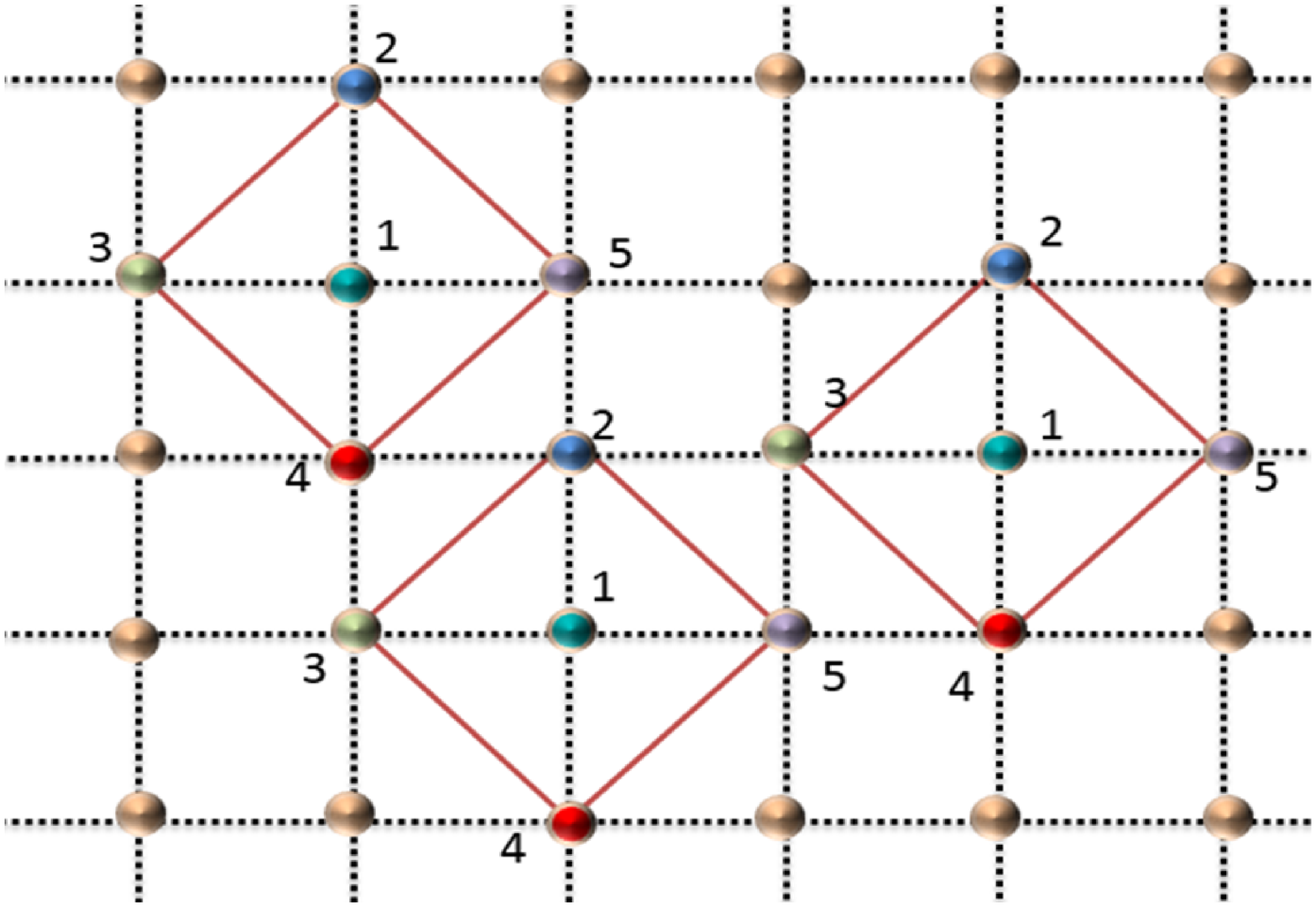} \put(0,65){(a)} \end{overpic}
		\hspace{0.5cm}
		\begin{overpic}[height=8cm]{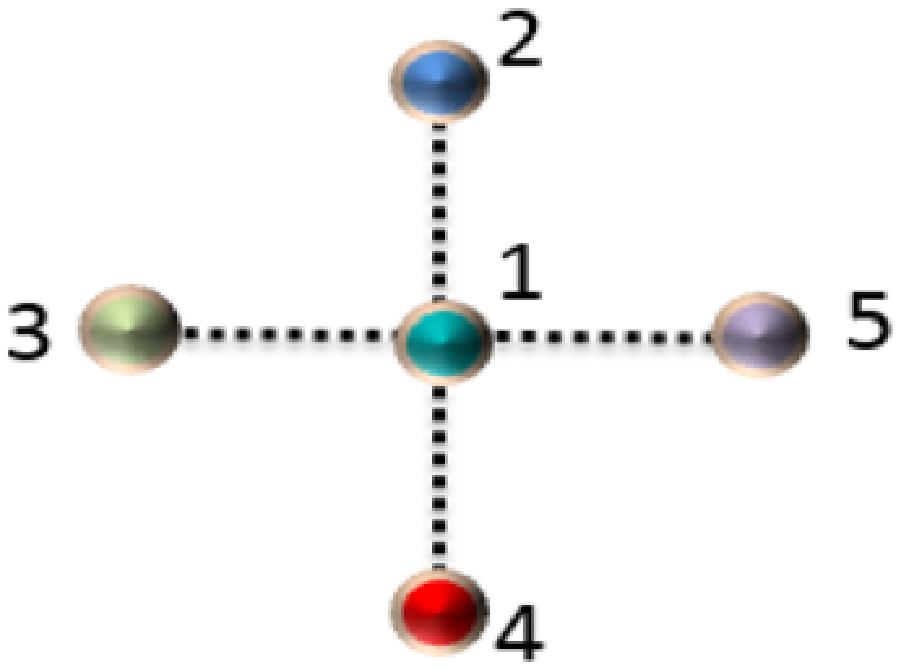} \put(20,55){(b)} \end{overpic}
		
	} \caption{(a) A two-dimensional square lattice with five-site blocking. (b) The sketch of one basic block.  }
	\label{fig:1}
\end{figure}

Consider the Hamiltonian $H$ which is defined on a square lattice. In Kadanof's approach, the lattice is divided into blocks and the total Hamiltonian $H$ can be written as
\begin{equation}
H=H^B + H^{BB},
\end{equation}
where $H^B$ is the Hamiltonian of all individual blocks and $H^{BB}$ is the interaction Hamiltonian of blocks. For example a five-site blocking can be applied to a square lattice, as it can be seen in \autoref{fig:1}. \\

Each block is then treated individually to build the projective operator onto its lower energy subspace, namely, the subspace spanned with $|\psi_0\rangle$ and $|\psi_1\rangle$. 
Projecting $H$ onto the derived subspaces, maps it to an effective Hamiltonian which acts on the renormalized subspace \cite{QRGG2,QRG Jafari,QRG XXZ}, i.e.

\be\label{Heff}
H^{eff}=P_0^{\dagger}HP_0= P_0^{\dagger}H^BP_0+P_0^{\dagger}H^{BB}P_0,
\ee
with the projective operator
\be\label{P.O.}
P_0=\prod_{L}^{}(|\Uparrow\ra_L\la\psi_0|+|\Downarrow\ra_L\la\psi_1|),
\ee
where $|\Uparrow\ra_L$ and $|\Downarrow\ra_L$ are the renamed states of block $L$ which represent the \textit{effective site} degrees of freedom, and the product over $L$ runs over all blocks. 
Different block structures can be considered on a square lattice, but in the QRG method, one should consider a structure for which $H^{eff}$ has exactly the same form as the original $H$, and it differs only in coupling constants. For example, for the XY Hamiltonian which will be studied in the next section, the effective site is derived from five primary sites as shown in \autoref{fig:1} \cite{QRG XY}. That is while for the two dimensional Ising model of \autoref{Ising}, it is more convenient to map four sites to an effective site \cite{Q. E. 2D}.\\

After deriving the iterative relation between the renormalized coupling constant (namely $g'$) and the original one ($g$), the critical point can be obtained by finding the fixed point of that iterative relation (i.e. $g'=g$). More details can be found in \cite{QRG XY, Q. E. 2D}, and we will use the results of these papers in the next two sections, in order to investigate the behavior of quantum coherence near the critical point of $XY$ and Ising models.

\section{XY model}\label{XYMODEL}
The two dimensional XY model for a  square lattice is described by the Hamiltonian 

\be\label{hamiltonian02}
H(J,\g)=\frac{J}{4} \sum_{i,j}[(1+\g)(\sigma_{i,j}^x\sigma_{i+1,j}^x+\sigma_{i,j}^x\sigma_{i,j+1}^x)+(1-\g)(\sigma_{i,j}^y\sigma_{i+1,j}^y+\sigma_{i,j}^y\sigma_{i,j+1}^y)],
\ee
where $J$ is the exchange coupling, $\g$ is the anisotropy parameter, $\sigma_{i,j}^{\alpha}$ with $\alpha=\{x,y\}$ is the Pauli $\alpha$ operator for the spin in row number $i$ and column $j$ of the square lattice, and the summation over $i$ and $j$ spans the lattice horizontally and vertically  respectively. The model corresponds to the spin fluid phase for $\gamma \longrightarrow 0$ and is called the XX model, while it reduces to Ising-like model for $\gamma \longrightarrow \pm 1$. To apply the QRG method, we partition the lattice into blocks of five spins as depicted in \autoref{fig:1}. Hence the block Hamiltonian is 
\begin{equation}\label{HB2}
H^B=\frac{J}{4} \sum_{L}^{}[(1+\g)(\sigma_{L,1}^x\sigma_{L,2}^x+\sigma_{L,1}^x\sigma_{L,3}^x+\sigma_{L,1}^x\sigma_{L,4}^x+\sigma_{L,1}^x\sigma_{L,5}^x)+(1-\g)(\sigma_{L,1}^y\sigma_{L,2}^y+\sigma_{L,1}^y\sigma_{L,3}^y+\sigma_{L,1}^y\sigma_{L,4}^y+\sigma_{L,1}^y\sigma_{L,5}^y)],
\end{equation}\\
where $\sigma_{L,m}^{\alpha}$ is the Pauli $\alpha$ operator of spin number  $m$ of block $L$. The lowest eigen-energy of the single-block Hamiltonian is $E_0=-\frac{1}{2} J \sqrt{5+ 5\gamma ^2 + \alpha_1}$, and the two corresponding eigenvectors are

\be\label{psi0}
\begin{split}
	|\psi_0\ra =&\g_1(|\uparrow\uparrow\uparrow\uparrow\downarrow\ra+|\uparrow\uparrow\uparrow\downarrow\uparrow\ra+|\uparrow\uparrow\downarrow\uparrow\uparrow\ra+|\uparrow\downarrow\uparrow\uparrow\uparrow\ra)\\
	&+\g_2(|\uparrow\uparrow\downarrow\downarrow\downarrow\ra+|\uparrow\downarrow\uparrow\downarrow\downarrow\ra+|\uparrow\downarrow\downarrow\uparrow\downarrow\ra+|\uparrow\downarrow\downarrow\downarrow\uparrow\ra)\\
	&+\g_3|\downarrow\uparrow\uparrow\uparrow\uparrow\ra+\g_4(|\downarrow\uparrow\uparrow\downarrow\downarrow\ra+|\downarrow\uparrow\downarrow\uparrow\downarrow\ra+|\downarrow\uparrow\downarrow\downarrow\uparrow\ra\\
	&+|\downarrow\downarrow\uparrow\uparrow\downarrow\ra+|\downarrow\downarrow\uparrow\downarrow\uparrow\ra+|\downarrow\downarrow\downarrow\uparrow\uparrow\ra)+\g_5|\downarrow\downarrow\downarrow\downarrow\downarrow\ra,
\end{split}
\ee
and
\be\label{psi1}
\begin{split}
	|\psi_1\ra =&\g_6|\uparrow\uparrow\uparrow\uparrow\uparrow\ra+\g_7(|\uparrow\uparrow\uparrow\uparrow\downarrow\ra+|\uparrow\uparrow\downarrow\uparrow\downarrow\ra+|\uparrow\uparrow\downarrow\downarrow\uparrow\ra\\
	&+ |\uparrow\downarrow\uparrow\uparrow\downarrow\ra+|\uparrow\downarrow\uparrow\downarrow\uparrow\ra+|\uparrow\downarrow\downarrow\uparrow\uparrow\ra+\g_8|\downarrow\downarrow\downarrow\downarrow\uparrow\ra\\
	&+\g_9(|\downarrow\uparrow\uparrow\uparrow\downarrow\ra+|\downarrow\uparrow\uparrow\downarrow\uparrow\ra+|\downarrow\uparrow\downarrow\uparrow\uparrow\ra+|\downarrow\downarrow\uparrow\uparrow\uparrow\ra)\\
	&+ \g_{10}|\downarrow\uparrow\downarrow\downarrow\downarrow\ra+|\downarrow\downarrow\uparrow\downarrow\downarrow\ra+|\downarrow\downarrow\downarrow\uparrow\downarrow\ra+|\downarrow\downarrow\downarrow\downarrow\uparrow\ra),
\end{split}
\ee
in which $|\uparrow\ra$ and $|\downarrow\ra$ are eigenvectors of $\sigma^z$ Pauli operator with eigenvalues $+1$ and $-1$ respectively, and explicit expressions of constants $\gamma_i$ are presented in Appendix. The eigenvectors (\ref{psi0}) and (\ref{psi1}) are chosen such that $\langle \psi_0 | \sigma_{1}^z\sigma_{2}^z \sigma_{3}^z\sigma_{4}^z \sigma_{5}^z | \psi_0 \rangle =+1 $ and $\langle \psi_1 | \sigma_{1}^z\sigma_{2}^z \sigma_{3}^z\sigma_{4}^z \sigma_{5}^z | \psi_1 \rangle =-1 $, and hence each state belongs to one half of the total Hilbert space. $P_0^L=|\psi_0\ra_L\la\Uparrow|+|\psi_1\ra_L\la\Downarrow|$ is now used to map the original Hamiltonian to the effective one, with the renormalized constants, i.e. 
\be\label{hamiltonian2}
H(J^{'},\g{'})=\frac{J{'}}{4} \sum_{i,j}[(1+\g^{'})(\sigma_{i,j}^x\sigma_{i+1,j}^x+\sigma_{i,j}^x\sigma_{i,j+1}^x)+(1-\g^{'})(\sigma_{i,j}^y\sigma_{i+1,j}^y+\sigma_{i,j}^y\sigma_{i,j+1}^y)],
\ee
with
\be\label{J'}
\begin{split}
	J^{'}=&j(\g_{10}^2(9\g_4^2+6\g\g_4\g_5+\g_5^2)+9\g_2^2\g_7^2+\g_1^2(\g_6^2+6\g\g_6\g_7+9\g_7^2)+6\g\g_2^2\g_7\g_8+\g_2^2\g_8^2+6\g\g_2\g_3\g_7\g_9+18\g_2\g_4\g_7\g_9\\
	&+2\g_2\g_3\g_8\g_9+6\g\g_2\g_4\g_8\g_9+\g_3^2\g_9^2+6\g\g_3\g_4\g_9^2+9\g_4^2\g_9^2+2\g_1\{\g_2[\,3\g_7(3\g\g_7+\g_8)+\g_6(3\g_7+\g\g_8)]\,\\
	&+(\g\g_3\g_6+3\g_4\g_6+3\g_3\g_7+9\g\g_4\g_7)\g_9\}+2\g_{10}\{\g_1(\g_5\g_6+9\g_4\g_7)+\g[\,9\g_2\g_4\g_7+3\g_1(\g_4\g_6+\g_5\g_7)\\
	&+\g_2\g_5\g_8+9\g_4^2\g_9]\,+3[\,\g_2(\g_5\g_7+\g_4\g_8)+\g_4(\g_3+\g_5)\g_9]\,\})
\end{split}
\ee
and

\be\label{gamma'}
\begin{split}
	\g^{'}=&[\,2(3\g_{10}\g_4+3\g_1\g_7+\g_2\g_8+\g_3\g_9)(\g_{10}\g_5+\g_1\g_6+3\g_2\g_7+3\g_4\g_9)+\g(\g_{10}^2(9\g_4^2+\g_5^2)+9\g_2^2\g_7^2+\g_1^2(\g_6^2+9\g_7^2)\\
	&+\g_2^2\g_8^2+18\g_2\g_4\g_7\g_9+2\g_2\g_3\g_8\g_9+\g_3^2\g_9^2+9\g_4^2\g_9^2+6\g_1[\,\g_2\g_7(\g_6+\g_8)+(\g_4\g_6+\g_3\g_7)\g_9]\,\\
	&+2\g_{10}\{\g_1(\g_5\g_6+9\g_4\g_7)+3[\,\g_2(\g_5\g_7+\g_4\g_8)+\g_4(\g_3+\g_5)\g_9]\,\})]\,/[\,\g_{10}^2(9\g_4^2+6\g\g_4\g_5+\g_5^2)+9\g_2^2\g_7^2\\
	&+\g_1^2(\g_6^2+6\g\g_6\g_7+9\g_7^2)+6\g\g_2^2\g_7\g_8+\g_2^2\g_8^2+6\g\g_2\g_3\g_7\g_9+18\g_2\g_4\g_7\g_9+2\g_2\g_3\g_8\g_9+\g_3^2\g_9^2\\
	&+6\g\g_3\g_4\g_9^2+9\g_4^2\g_9^2+2\g_1\{\g_2[\,3\g_7(3\g\g_7+\g_8)+\g_6(3\g_7+\g\g_8)]\,+(\g\g_3\g_6+3\g_4\g_6+3\g_3\g_7+9\g\g_4\g_7)\g_9\}\\
	&+2\g_{10}(\g_1(\g_5\g_6+9\g_4\g_7)+\g[\,9\g_2\g_4\g_7+3\g_1(\g_4\g_6+\g_5\g_7)+\g_2\g_5\g_8+9\g_4^2\g_9+\g_3\g_5\g_9]\,\\
	&+3[\,\g_2(\g_5\g_7+\g_4\g_8)+\g_4(\g_3+\g_5)\g_9]\,)]\,.
\end{split}
\ee

Having the iteration relations, the next step is to choose a quantum coherence measure to quantity the coherence of intended states. Here we have used the $l_1$ norm of coherence which, for an arbitrary density matrix $\rho$, is the summation of all off-diagonal elements, 

\be\label{norml_1}
QC_{l_1}(\rho)=\sum_{\substack{
		\\ i\neq j}}\lvert\rho_{i,j}\rvert.
\ee

\begin{figure}[H]
	\includegraphics[width=11cm,height=8cm]{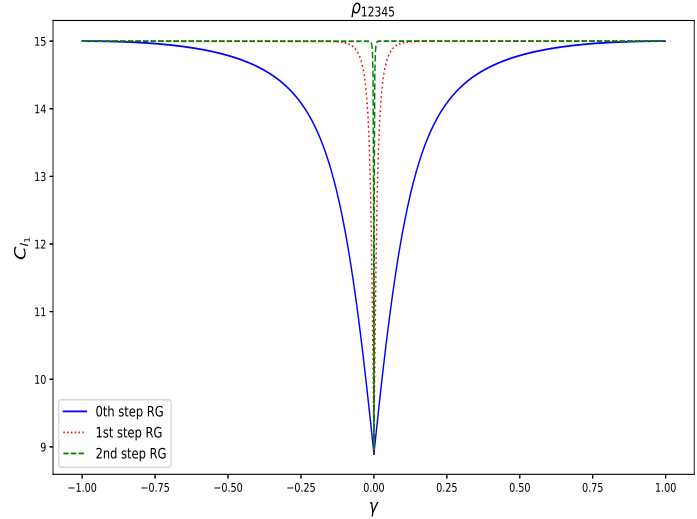}
	\centering
	\vspace{-0.5cm}
	\caption{(Color Online) Quantum coherence of five-site ground state $\rho_0$ of the XY model, as a function of $\g$  and for $J=1$. }
	\label{CoherenceXY}
\end{figure}

\begin{figure}[H]
	\includegraphics[width=10.5cm,height=7cm]{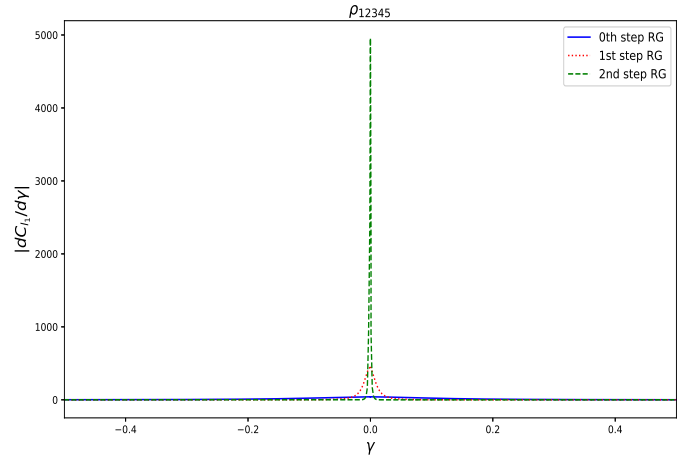}
	\centering
	\vspace{-0.5cm}
	\caption{(Color Online) The first derivative of quantum coherence of $\rho_0$, for XY model and as a function of $\g$, we have considered $J=1$.}
	\label{dXY}
\end{figure}

To investigate the behavior of quantum coherence, we consider the ground state density matrix $\rho_{0}=|\psi_0\ra\la\psi_0|$ and plot the quantum coherence and its first derivative as functions of $\gamma$, for different renormalization steps (see \autoref{CoherenceXY} and \autoref{dXY}). Note that a large system with $N=5^{n+1}$ spins can be depicted by a block with $5$ effective sites for $n-$th renormalization step.\\

It is clear in \autoref{CoherenceXY} that $\gamma=0$ is a candidate of critical point, since the curves become sharper in $\gamma=0$ while the renormalization iteration steps increase  (or alternatively, the number of spins increases). This can also be seen in \autoref{dXY}, where the non-analytic behavior of $\left| \frac{dC_{l_1}}{d\g} \right|$ is more evident. These observations, show that quantum coherence can be regarded as a QPT-detector for the XY model. In fact, after three steps, quantum coherence of $\rho_0$ attains two fixed values, the maximal possible value, $15$, for $\g \neq 0$ and a fixed non-maximal value for $\g= 0$ \footnote{Note that, here the maximal possible value is $15$ since $|\psi_0\rangle$ belongs to one half of the Hilbert space.}. These values clearly show the difference between symmetries of $\g=0$ and $\g \neq 0$. The points $\gamma=1$ and $\gamma=-1$, correspond to two Ising phases in $x$ and $y$ directions and then the state $|\psi_0\rangle$ reduces to   $\frac{1}{\sqrt{2}} \left( |x_- x_+ x_+ x_+ x_+ \rangle + |x_+ x_- x_- x_- x_- \rangle \right)$ or $\frac{1}{\sqrt{2}} \left( |y_- y_+ y_+ y_+ y_+ \rangle+  |y_+ y_- y_- y_- y_- \rangle \right)$ respectively which clearly show the long range order of the system. In contrary, the non-maximal value of coherence at $\gamma =0$ confirms that the system is coherent with no long range order, due to the presence of quantum fluctuations. 

\begin{figure}[H]
	\vspace{-0.5cm}
	\includegraphics[width=13cm,height=10.5cm]{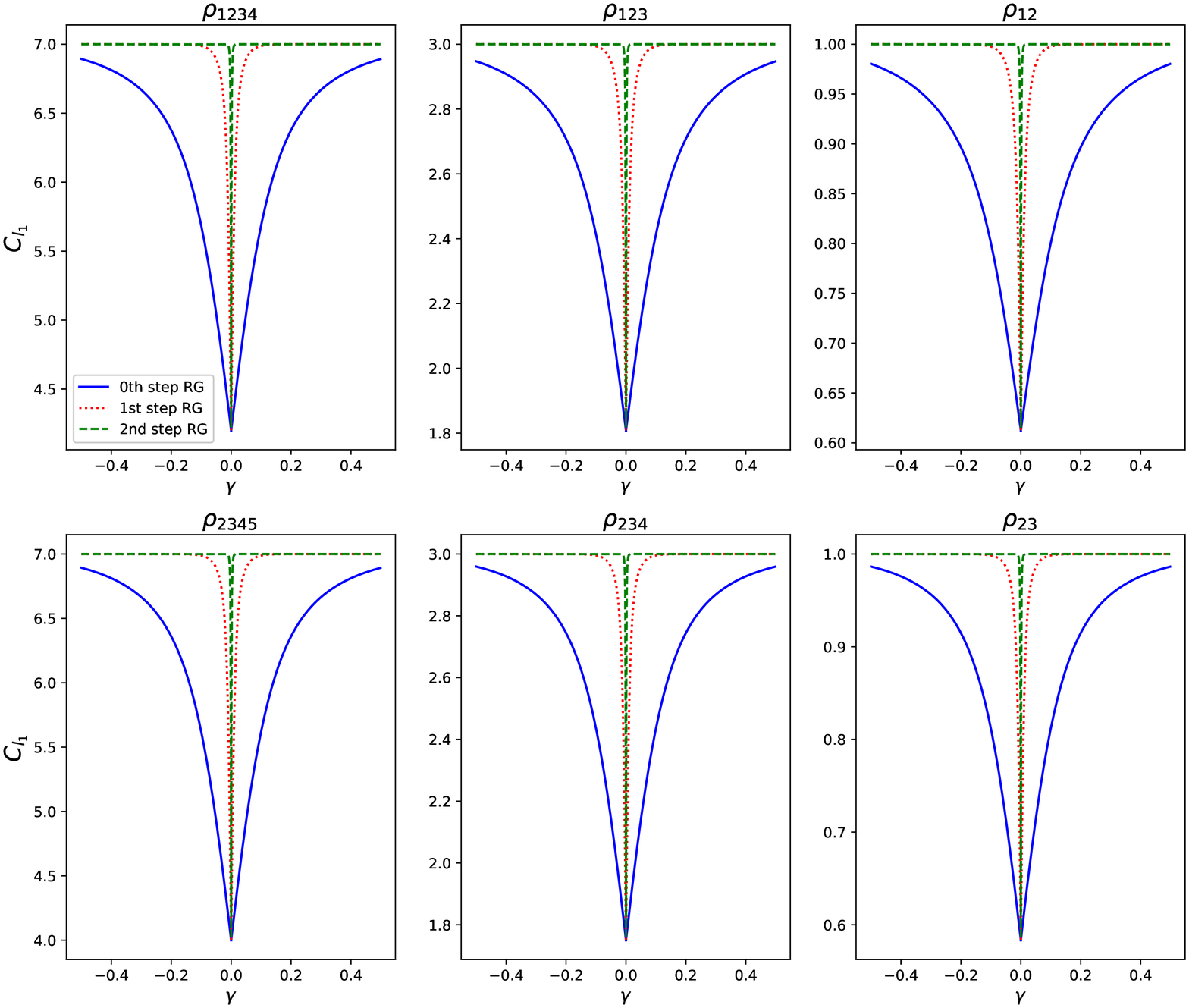}
	\centering
	\vspace{-0.5cm}
	\caption{(Color Online) First three renormalization steps for quantum coherence of different reduced density matrices obtained from $\rho_0$ of XY model. }
	\label{XYpartialT}
\end{figure}

\begin{figure}[H]
	\vspace{-0.5cm}
	\includegraphics[width=13cm,height=10.5cm]{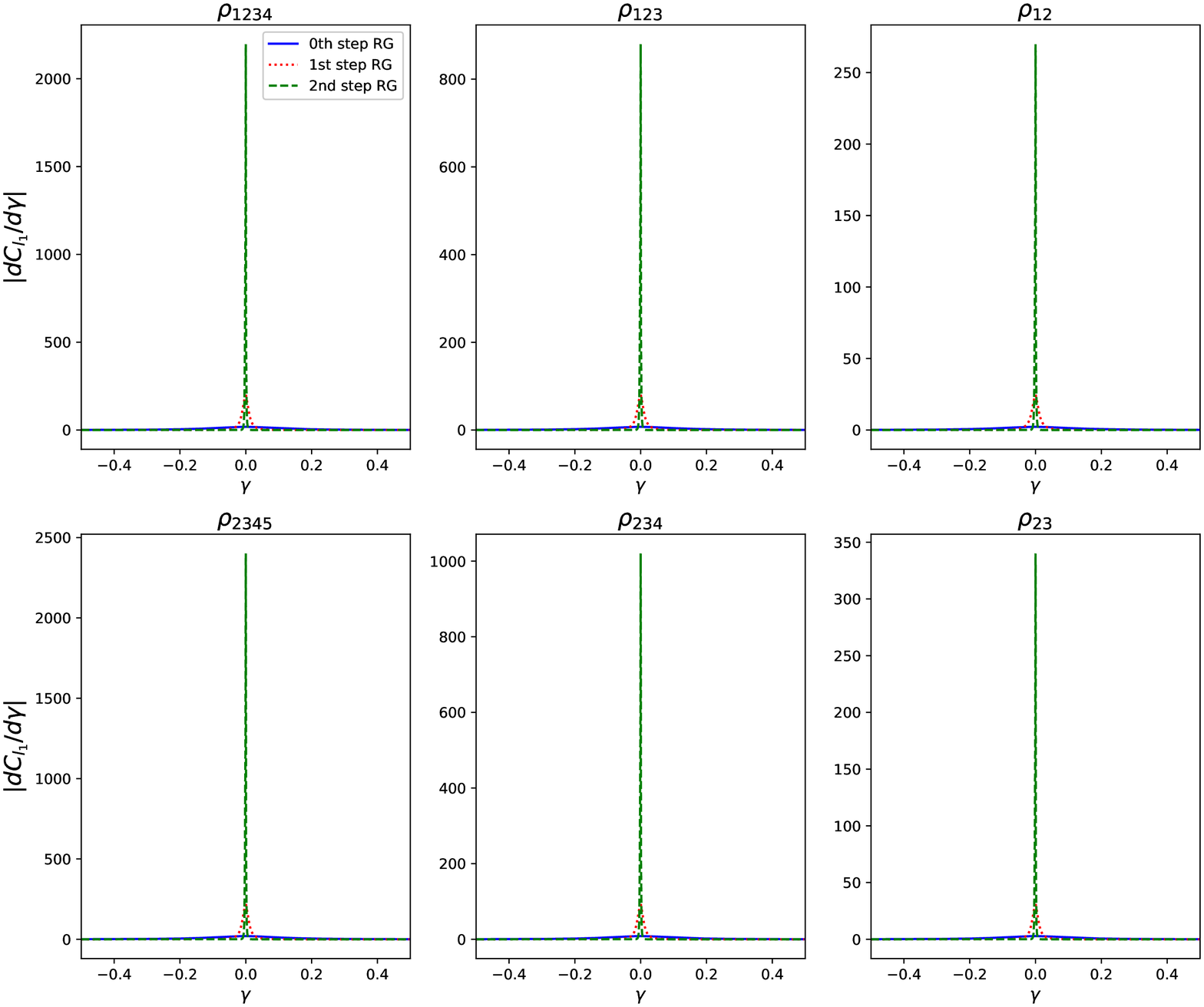}
	\centering
	\vspace{-0.5cm}
	\caption{(Color Online) The first derivative of quantum coherence of  different reduced density matrices of $\rho_0$, for the first three renormalization steps.}
	\label{XYpartialTD}
\end{figure}

A similar singular behavior is also seen for quantum coherence of marginal density matrices $\rho_0$, namely the states $\rho_{1234}=tr_5(\rho_0)$,  $\rho_{123}$, $\rho_{12}$ and etc. The quantum coherences of marginal states and their derivatives are plotted in \autoref{XYpartialT} and \autoref{XYpartialTD} respectively. It is seen that, for the general behavior of coherence, there is no difference between the partitions which contain the central site $1$ and the ones which do not. For all marginal states, after three renormalization steps, the quantum coherence tends to a two-value function, the maximal possible value for $\gamma \neq 0$ and a non-maximal value for $\gamma =0$.
Note that the coherences of two-site marginal states ($\rho_{12}$ or $\rho_{23}$) have also singular behavior and show the critical point, it is while they are easy to  calculate. As a matter of fact, the global properties of the system are effectively included in the quantum coherence of two sites. It is valuable to notice that calculations of quantum coherence are computationally more convenient than multi-partite or two-partite entanglements \cite{QRG XY}.


\begin{figure}[H]
	\includegraphics[width=11cm,height=8cm]{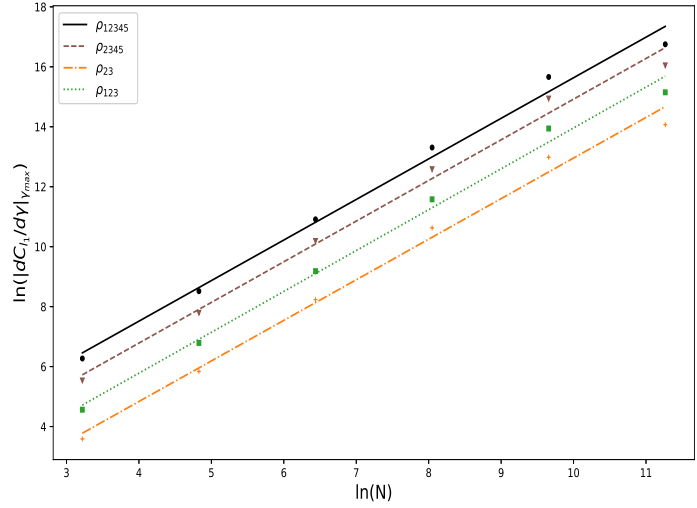}
	\centering
	\caption{(Color Online) The scaling behavior of $Ln|\frac{dC}{d\g}|_{\g=0}$ in terms of $Ln(N)$.}
	\label{XYLn}
\end{figure}

To investigate the scaling behavior of $f:=\big|\frac{dC}{d\g}\big|_{max}$, with respect to the system size $N$, in \autoref{XYLn}, we plot $Ln(f)$ as a function of $Ln(N)$ for different subsystem density matrices. All curves show similar linear behavior and hence $\big|\frac{dC}{d\g}\big|_{max} \sim N^\theta$ with the coherence exponent $\theta=1.36$, which is very close to entanglement exponent $1.35$ that is obtained by entanglement analyses \cite{QRG XY}. In the next section we will consider the Ising model with the transverse field and we show that two critical exponents of the model can be obtained by coherence analysis. 

\section{Ising model with transverse field}\label{Ising}
The other system which we consider is a spin-$\frac{1}{2}$ transverse-field Ising system with  Hamiltonian
\begin{equation}\label{hamiltonian1}
H=-J\sum_{\left<i,j\right>}\sigma_i^z\sigma_j^z-h\sum_{i}\sigma_i^x,
\end{equation}
on a square lattice. Here $J>0$ is the ferromagnetic exchange coupling, $\sigma_i^\alpha$ ($\alpha=x,z$) are Pauli operators on site $i$, $h \ge 0$ is the transverse magnetic field, and the summation runs over all nearest neighbors. The constant $J$ can be factorized and the normalized transverse field will be defined as $g =\frac{h}{J}$.\\

In order to investigate the behavior of quantum coherence in the vicinity of the critical point of this model, we use the results of the conventional RG method which is applied to 2D Ising model \cite{Q. E. 2D}. According to that scheme, the lattice is partitioned into blocks of size $4$, and each block is then divided in horizontal and vertical directions, which will be mapped to the effective sites. To preserve the symmetry of the system, the order of both renormalization directions should be equivalent, and from there the renormalized value of couplings are derived.
After implementing this idea of RG, the effective normalized transverse field $g'=\frac{h'}{J'}$ of the 2D Ising model is obtained to be \cite{Q. E. 2D}

\begin{equation}\label{g'}
g^{'}=\frac{g^4((1+g^2)^3(4+4g^2+2g^4+g^6))^{\frac{1}{4}}}{(2+g^2)\sqrt{8+8g^2+3g^4+g^6}}.
\end{equation}

\begin{figure}[H]
	\includegraphics[width=11cm,height=7.5cm]{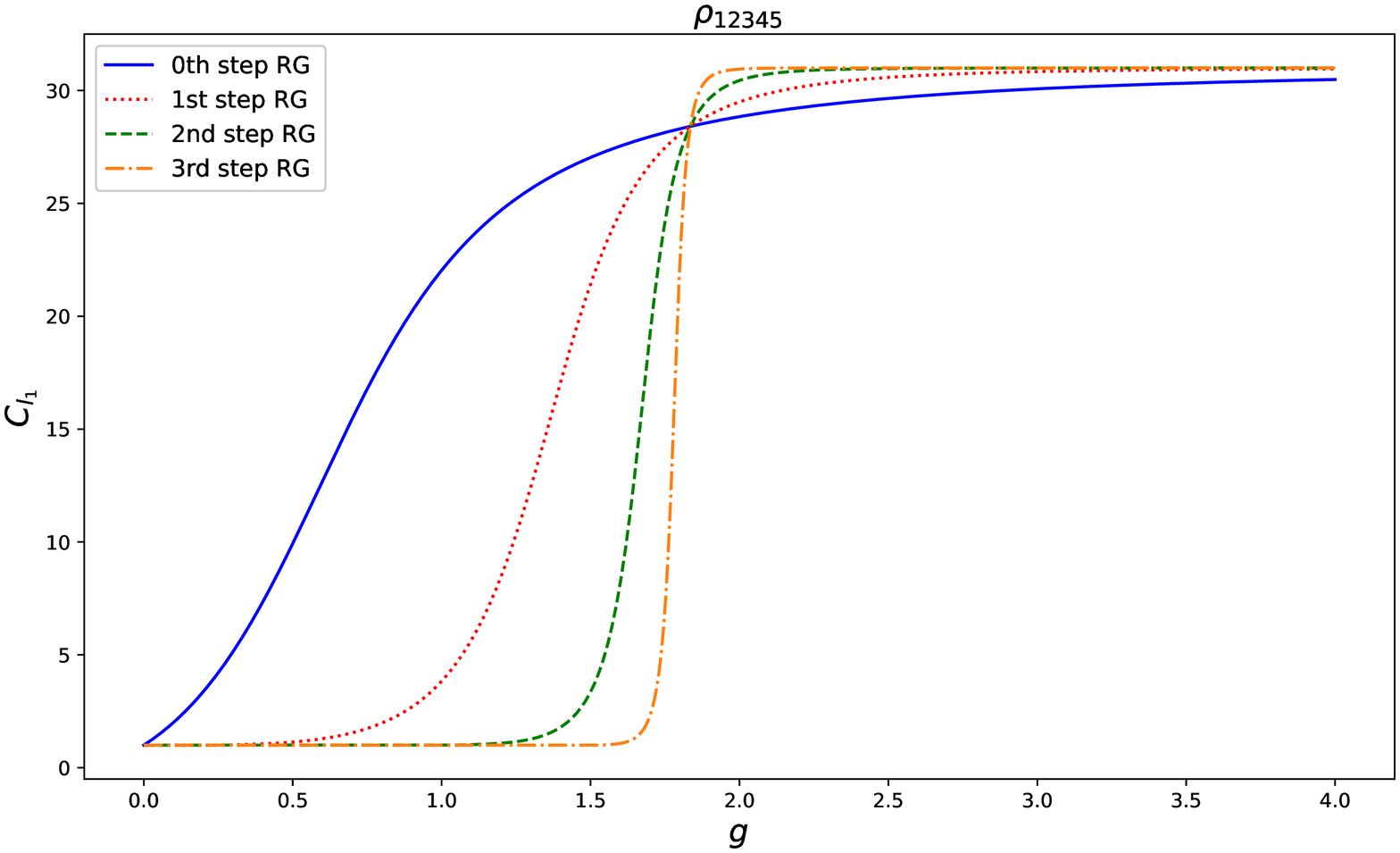}  
	\centering	
	\vspace{-0.5 cm}
	 \caption{(Color Online) The $l_1$-norm of coherence for the ground state of \autoref{HBIsing}, for different RG steps.}
	\label{fig:2}
\end{figure}

As the basic cluster to study the coherence, we again consider the 5-site block of \autoref{fig:1}, and hence one-block Hamiltonian is
\begin{equation}\label{HBIsing}
H^b= J\big[-(\sigma_{L,1}^z\sigma_{L,2}^z+\sigma_{L,1}^z\sigma_{L,3}^z+\sigma_{L,1}^z\sigma_{L,4}^z+\sigma_{L,1}^z\sigma_{L,5}^z)-g (\sigma_{L,1}^x+\sigma_{L,2}^x+\sigma_{L,3}^x+\sigma_{L,4}^x+\sigma_{L,5}^x) \big],
\end{equation}
and the effective block of $n-$th RG step is equivalent to a lattice of size $N=5\times 4^n$.
We numerically calculate the ground state of \autoref{HBIsing}, $\rho_0$, and plot its $l_1$ norm of quantum coherence for different values of $g$ and different RG steps in \autoref{fig:2}. 

\begin{figure} [H]
 	\includegraphics[width=13cm,height=9cm]{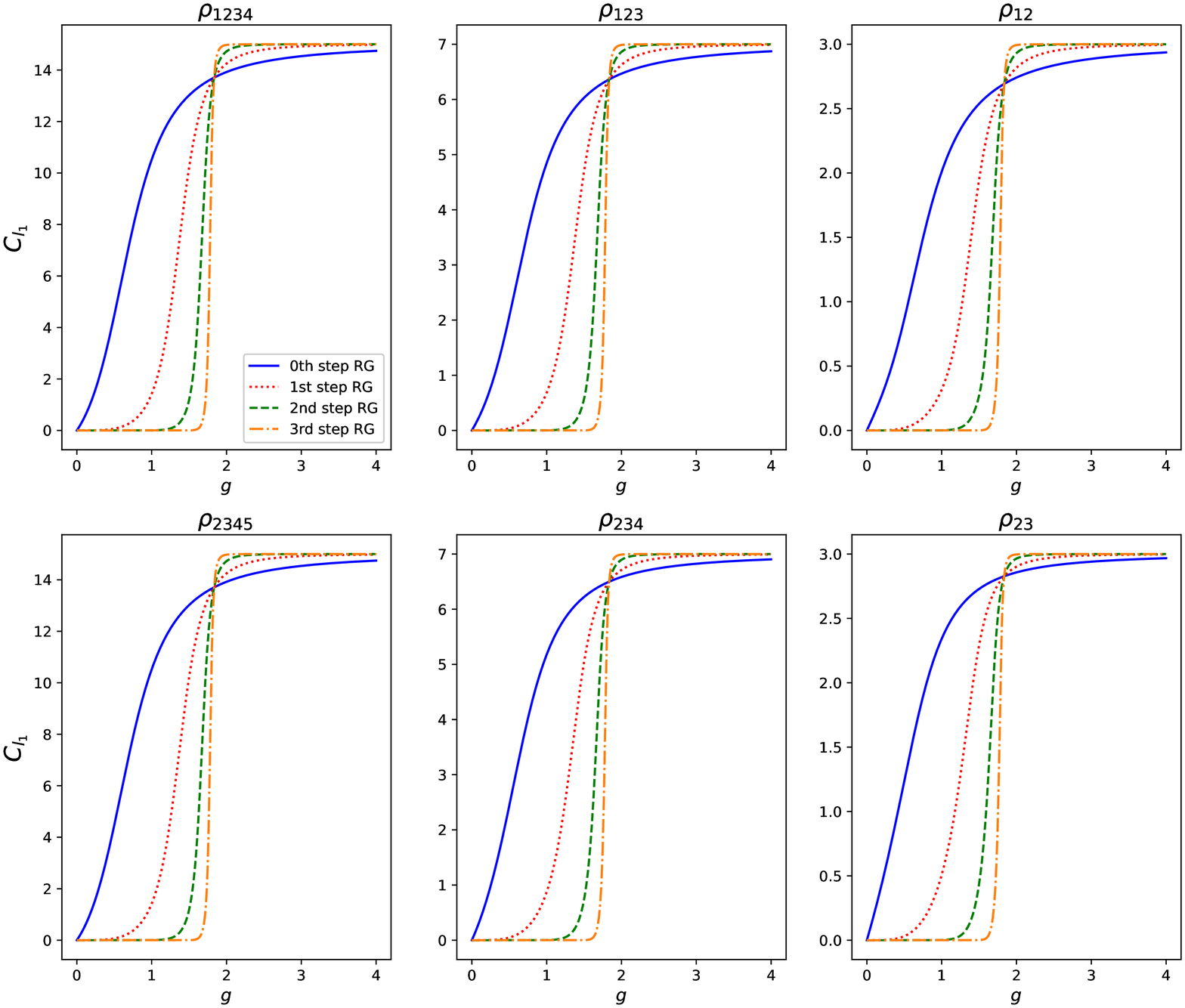}
 	\centering
 	\vspace{-0.5 cm}
 	\caption{(Color Online) Quantum coherence of various marginal states of the Ising system ground state, for different QRG iteration steps.  }
 	\label{partial trace}
 \end{figure}

It is evident in \autoref{fig:2} that, for the first two iterations, which correspond to small lattice sizes, the quantum coherence of $\rho_0$ increases gradually. On the other hand, for higher iterations which represent larger lattices, the quantum coherence increases suddenly when the parameter $g$ tends to its critical value. Note that the ground state of \autoref{HBIsing} is dually degenerate for $g=0$ and we have chosen the state $\frac{1}{\sqrt{2}}\left(|\uparrow \uparrow \uparrow \uparrow \uparrow  \rangle +|\downarrow \downarrow \downarrow \downarrow \downarrow \rangle \right)$ to avoid mathematical discontinuity in $g=0$. In fact minimal value $1$ of coherence, corresponds this state and the maximal value stands for the state $|x_+ x_+ x_+ x_+ x_+ \rangle$ which shows the ferromagnetic ordered phase. To inquire the behavior of coherence for reduced density matrices, we did the same calculations for the marginal states, and the results are shown in \autoref{partial trace}. The figure clearly shows that, in the vicinity of critical point, the general behavior of coherence is the same for all marginal states. Due to the non-analyticity which is seen in higher RG steps, the quantum coherence of $\rho_0$ and its marginal states can be regarded as QPT-detectors.\\

\begin{figure} [H]
	\includegraphics[width=9cm,height=6.5cm]{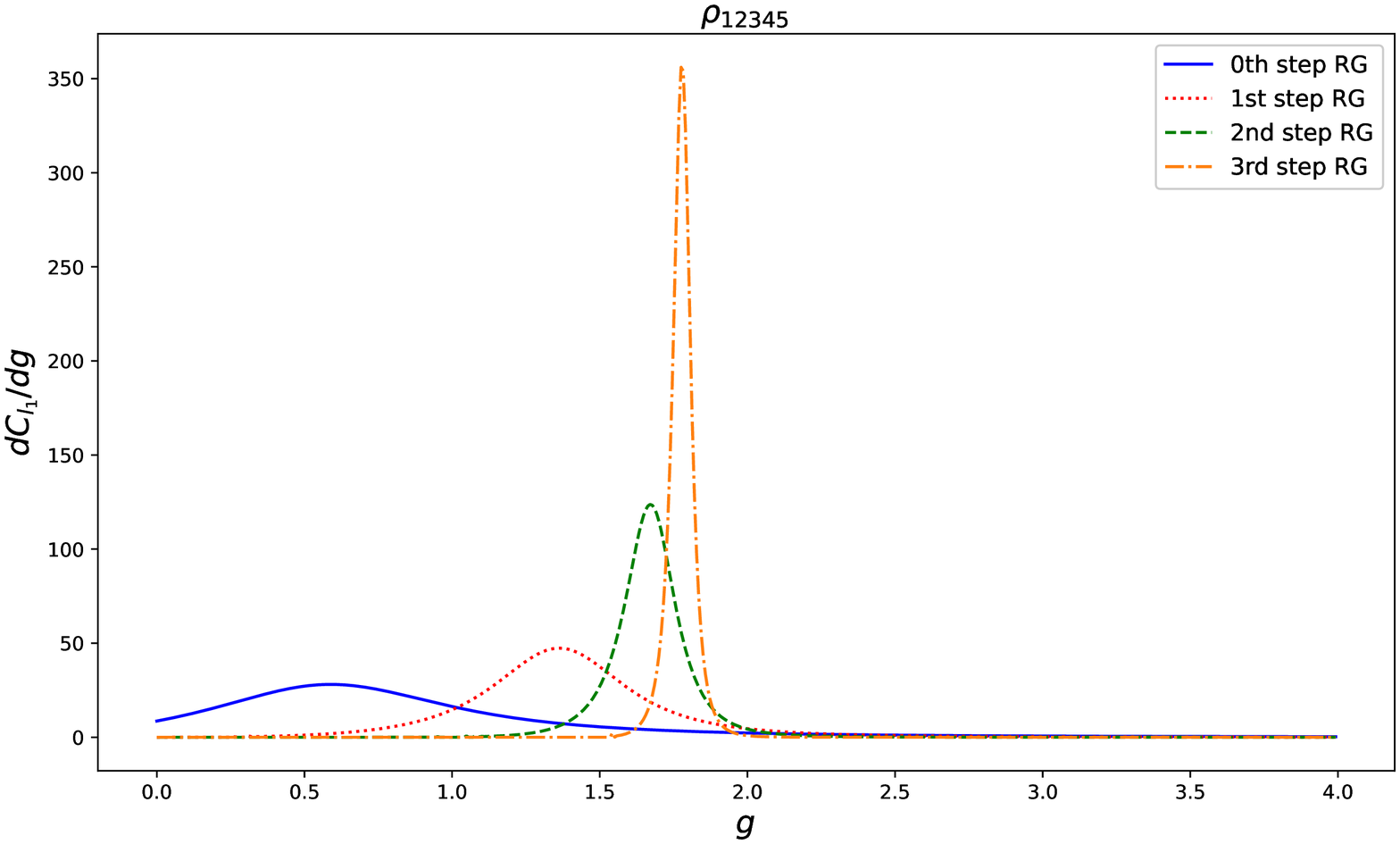} 
	\centering	
	\vspace{-0.5 cm}
	\caption{(Color Online) The first derivative of quantum coherence for different QRG iterations steps of the ground state of Ising model.}
	\label{fig:3}
\end{figure}

\begin{figure}[H]
 	\includegraphics[width=13.5cm,height=9cm]{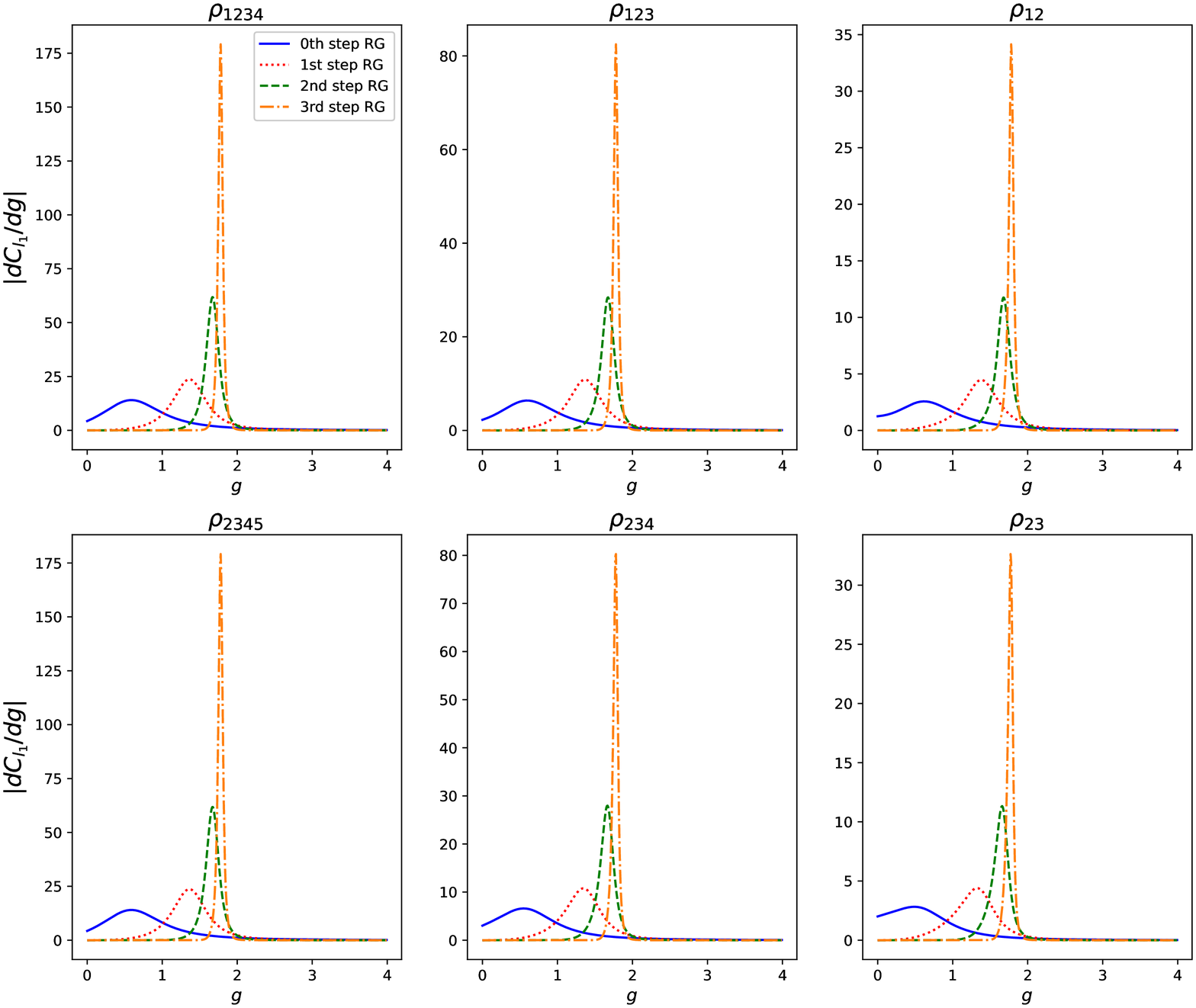}
 	\centering
 	\vspace{-0.5 cm}
 	\caption{(Color Online) First derivative of quantum coherence for different QRG steps and for some reduced states of the ground state of Ising model. }
 	\label{divert partial trace}
 \end{figure}

To provide more precise results, we plot the first derivative of coherence as a function of $g$, for $\rho_0$ and its corresponding reduced density matrices, and for different RG iterations. The results are presented in \autoref{fig:3} and \autoref{divert partial trace}. It is seen that  there is a local maximum in
$\big|\frac{dC_{l_1}}{dg}\big|$ and the position of maximum tends to the critical point $g_c=1.858$ by increasing the number of RG steps. Note that the value of $g_c$ is obtained to be the same for all figures.\\

\begin{figure}[H]
	\includegraphics[width=11cm,height=8cm]{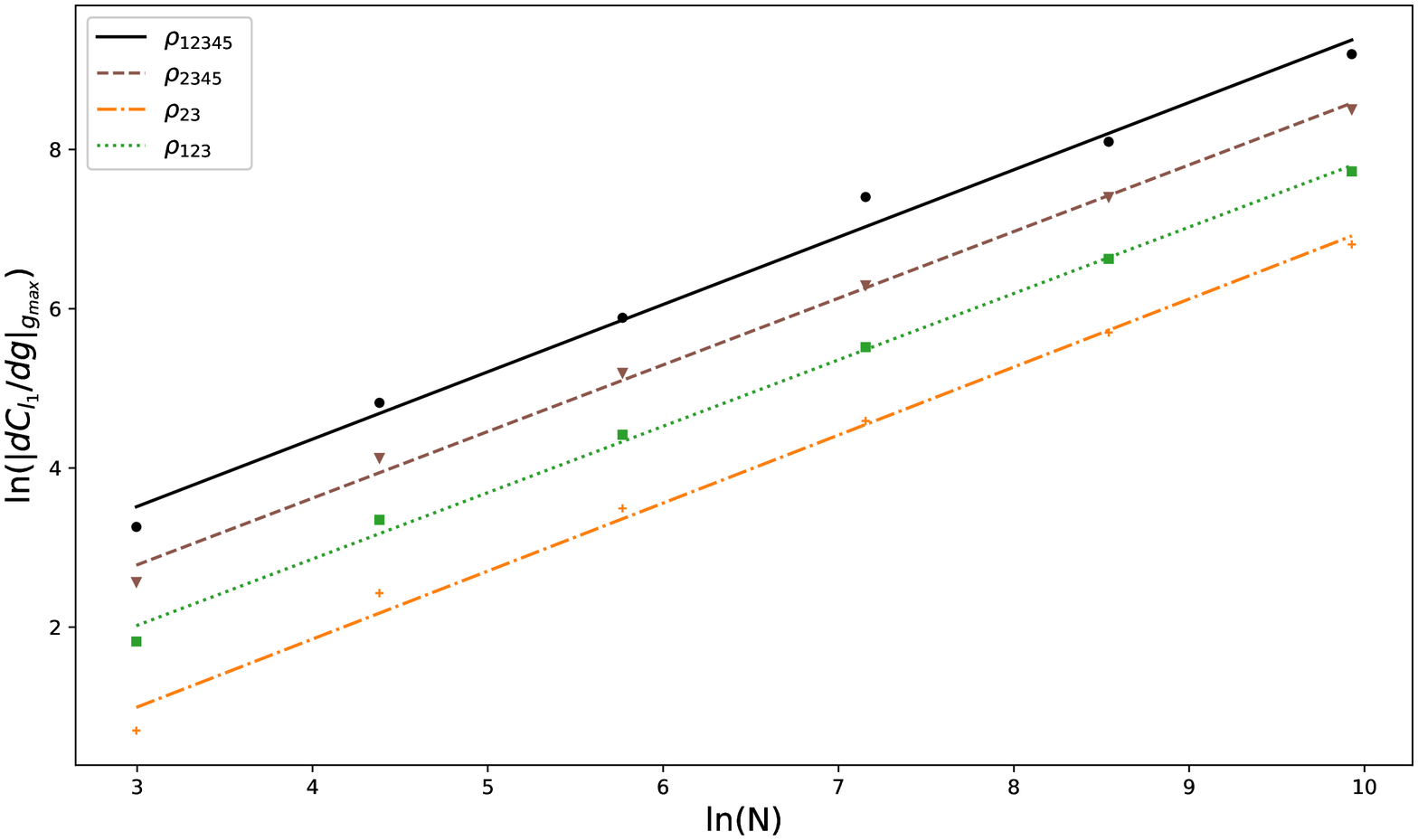}
	\centering
	\caption{(Color Online) The scaling behavior of $Ln\left(|\frac{dC}{dg}|_{g_{max}}\right)$ versus system size $Ln(N)$ for the ground state of the Ising model.}
	\label{lnIsing}
\end{figure}

To further elucidate the effect of system size on scaling behavior of the  maximum of $\big|\frac{dC_{l_1}}{dg}\big|$, we plot  the logarithm of $\big|\frac{dC}{dg}\big|_{g_{max}}$ as a function of the logarithm of system size, ln($N$) in \autoref{lnIsing}. As we discussed earlier, for the applied RG method and the basic block that we have chosen, $N=5 \times 4^{n}$ and the scaling low is derived to be $\big|\frac{dC}{dg}\big|_{g_{max}} \sim {N^{\theta}}$ with $\theta=0.84$ . 
It is shown in \cite{QRG XY} and \cite{Q. E. 2D} that, for a 2D lattice, the entanglement exponent $\theta$ is related to correlation length exponent $\nu$, by $\theta=1/2\nu$. Following the same steps as \cite{QRG XY}, we see that the same relation holds for the coherence exponent $\theta$, and thence the divergent behavior of correlation length of the 2D Ising model in the vicinity of critical point reads $\xi \sim (g-g_c) ^{-\nu}$ with $\nu=0.59$, which is close to the result $0.63$, obtained in \cite{QPT Ising}.\\
\\

It is noteworthy that, the critical point $g_c$ can also be easily obtained by finding the non-trivial fixed point of the iteration relation, i.e. $g^{'}=g$. The value is obtained to be $g_c=1.835$ which is near the singularity point of coherence $1.858$, but is different from the actual accurate quantum phase transition point through Monte Carlo methods $g_c=3.04$ \cite{Q. E. 2D}. Equation (\ref{g'}) can also be used to calculate the correlation length exponent $\nu$ as 

\be\label{exponent}
\n^{-1} = \log_2{\frac{dg^{'}}{dg}}\rvert_{g=g_c},
\ee
and the derived exponent $\n=0.63$, is very close to exact result of  numerical methods  $\n=0.6211$ \cite{Q. E. 2D}. The obtained values again show that applying  the renormalization method to quantum coherence can illustrate the long-distance critical properties which are independent of the system details although this method cannot obtain the critical point exactly.\\

It is also worth to mention that all the numerical calculations has been carried out by using the open-source QUTIP package \cite{QUTIP1,QUTIP2}.

\section{Conclusion}\label{conclusion}
We studied the behavior of $l_1$-norm quantum coherence near the critical point of two dimensional XY model and Ising model in the transverse magnetic field, by using the quantum renormalization method. We show that the coherence of five effective sites and also all its marginal states can be used to detect the QPT of these two models, this is while the calculations of quantum coherence is computationally much easier than entanglement and quantum discord. The results indicate that, near the critical point, the first derivative of quantum coherence is not continuous and hence quantum coherence is regarded as a quantity which detects quantum phase transition of these models. We also analyzed  the scaling behavior for the maximum of coherence derivative, with system size. Therein we found the critical exponents for coherence of both models and the results are in close agreement with the ones obtained by entanglement analysis. For the XY model, the quantum coherence tends to a two value function for higher RG steps, a non-maximal value for $\gamma=0$ and the maximal value for $\g\neq 0$, the values correspond to spin-fluid and Ising phases respectively. For the higher RG steps of the Ising model, we have also a two value function, with the minimum value for the disordered phase and the maximum possible value for the ferromagnetic phase. The connection between the behavior of quantum coherence and QPT of other two dimensional methods is still an open problem which can be considered for future works.

\appendix{}

\section*{Appendix}
\label{section:appendix}

The explicit expressions of $\g_{i}$'s of \autoref{psi0} and \autoref{psi1} are given by
\begin{gather*}
\g_1=-\frac{(-1+\a_1+\g^2)\sqrt{(5+ \alpha \a_1+5\g^2)}}{4\sqrt{2\a_2}},
\quad    
\g_2=-\frac{3\sqrt{\frac{\g^4(5+\a_1+5\g^2)}{\a_2}}}{{2\g\sqrt{2}}},
\quad 
\g_3=\frac{(-1+\a_1+\g^2)}{\sqrt{2\a_2}},
\quad
\g_4=\frac{\g(5+\a_1+\g^2)}{2\sqrt{2\a_2}}
\\
\g_5=\frac{3\sqrt{2}\g^2}{\sqrt{\a_2}} 
\quad
\g_6=\frac{\sqrt{\frac{\g^2(5+\a_1+5\g^2)}{1+\a_1+34\g^2-\a_1\g^2+\g^4}} (-2-2\a_1+17\g^2-3\a_1\g^2+3\g^4)}{4(3+2\g^2+3\g^4)}\\
\g_7=-\frac{\sqrt{\frac{\g^2(5+\a_1+5\g^2)}{1+\a_1+34\g^2-\a_1\g^2+\g^4}}(1+\a_1-\g^2+6g^4)}{4\g(3+2\g^2+3\g^4)}, 
\quad 
\g_8=-\frac{3\sqrt{\frac{\g^2(5+\a_1+\g^2)}{(1+\a_1+34\g^2-\a_1\g^2+\g^4}}(5-\a_1+5\g^2)}{4(3+2\g^2+3\g^4)}, \\
\g_9=\frac{(1+\a_1-\g^2)}{4\g\sqrt{(34-\a_1+\frac{1+\a_1}{\g^2}+\g^2)}},
\quad 
\g_{10}=\frac{3}{2\sqrt{(34-\a_1+\frac{1+\a_1}{\g^2}+\g^2)}},
\end{gather*} 
where 
\begin{gather*}
\alpha_1=\sqrt{1+34\g^2+\g^4},
\quad
\alpha_2=2-2\a_1+71\g^2+17\a_1\g^2+104\g^4+3\a_1\g^4+3\g^6.
\end{gather*}

{}


\begin{thebibliography}{}
	\bibitem{InfComp}M. A. Nielsen, I. L. Chuang, Quantum Computation and Quantum Communication, Cambridge University Press, Cambridge (2000).
	\bibitem{SpinModel1} G. Karpat, B. Cakmak, and F. F. Fanchini, Phys. Rev. B \textbf{90}, 104431 (2014).
	\bibitem{SpinModel2} B. Cakmak, G. Karpat, and F. F. Fanchini, Entropy \textbf{17}, 790 (2015).
	\bibitem{SpinModel3} A. L. Malvezzi, G. Karpat, B. Cakmak, F. F. Fanchini, T.
	Debarba, and R. O. Vianna, Phys. Rev. B \textbf{93}, 184428 (2016).
	\bibitem{SSPhys1} B. Deveaud-Pledran, A. Quattropani, P. Schwendimann, Quantum Coherence in Solid State Systems, Vol. \textbf{171} of Proceedings of the International School of Physics Enrico Fermi (IOS Press, Amsterdam, 2009).
	\bibitem{SSPhys2} Ch. Li, N. Lambert, Y. Chen, G. Chen, and F. Nori, Sci. Rep. \textbf{2},	885 (2012).
	 \bibitem{Condensed matter}H. Vazquez, R. Skouta, S. Schneebeli, M. Kamenetska, R. Breslow, L. Venkatara-man, M. Hybertsen, Nat. Nanotechnol. \textbf{7} 663 (2012).		
	\bibitem{Quantum Optics}M.O. Scully, M.S. Zubairy, Quantum Optics, Cambridge University, Cambridge (1997).
	\bibitem{metrology}V. Giovannetti, S. Lloyd, and L. Maccone, Nat. Photon. \textbf{5}, 222 (2011).
	\bibitem{biology}	N. Lambert, Y.-N. Chen, Y.-C. Cheng, C.-M. Li, G.-Y. Chen, and F. Nori, Nat. Phys. \textbf{9}, 10 (2013).
	\bibitem{Quantum Resource Theories1}	E. Chitambar and G. Gour, Rev. Mod. Phys. \textbf{91}, 025001 (2019).
	\bibitem{Quantum Resource Theories2}	A. Streltsov, S. Rana, P. Boes, J. Eisert, Phys. Rev. Lett. \textbf{119}, 140402 (2017.)
	\bibitem{Fidelity_Quantifying Coherence}	L.-H. Shao, Z. Xi, H. Fan, and Y. Li, Phys. Rev. A \textbf{91}, 042120 (2015).
	\bibitem{Intrinsic Randomness}X. Yuan, H. Zhou, Z. Cao, and X. Ma, Phys. Rev. A \textbf{92}, 022124 (2015).
	\bibitem{Measuring Concurrence}	X. Qi, T. Gao, and F. Yan, J. Phys. A: Math. Theor. \textbf{50}, 285301 (2017).
	\bibitem{Measuring with Entanglement}A. Streltsov, U. Singh, H. S. Dhar, M. N. Bera, and G. Adesso, Phys. Rev. Lett. \textbf{115}, 020403 (2015).
	\bibitem{Observable Measure}D. Girolami, Phys. Rev. Lett. \textbf{113}, 170401 (2014).
	\bibitem{Quantum Fisher}K. C. Tan, S. Choi, H. Kwon, and H. Jeong, Phys. Rev. A \textbf{97}, 052304 (2018).
	\bibitem{basis independent measure}C. Radhakrishnan, M. Parthasarathy, S. Jambulingam, and T. Byrnes, Phys. Rev. Lett. \textbf{116}, 150504 (2016).	
	 \bibitem{Quantifying Coherence}	T. Baumgratz, M. Cramer, and M. B. Plenio, Phys. Rev. Lett. \textbf{113}, 140401 (2014).	
	\bibitem{Phase Transitions} S. Sachdev, Quantum Phase Transitions, Second edition, Cambridge University Press, Cambridge; New York (2011).
	

	\bibitem{entanglement and QPT}F.-W. Ma, S.-X. Liu, and X.-M. Kong, Phys. Rev. A \textbf{83}, 062309 (2011).
	\bibitem{multi entanglement xxz model}T. Wang and J. Shi, Laser Phys. Lett. \textbf{16}, 065201 (2019).
	
	

	
	
	
	\bibitem{Multipartite Entanglement and QPT}	A. Montakhab and A. Asadian, Phys. Rev. A \textbf{82}, 062313 (2010).

	\bibitem{QPT nature}A. Osterloh, L. Amico, G. Falci, and R. Fazio, Nature \textbf{416}, 608 (2002)
	\bibitem{Q. Correlation}T. Werlang, C. Trippe, G. A. P. Ribeiro, and G. Rigolin, Phys. Rev. Lett. \textbf{105}, 095702 (2010).
	
	\bibitem{QC_Sci} YC. Li and HQ. Lin, Sci. Reports \textbf{6}, 26365 (2016).	
	
	\bibitem{Dynamical Behavior Ising}	M. Qin, L. Wang, M. He, and X. Wang, Physica A \textbf{540}, 122944 (2020) 
   
    \bibitem{Dynamics of XY}M. Qin, Z. Ren, and X. Zhang, Phys. Rev. A \textbf{98}, 012303 (2018). 
    
    \bibitem{QPT Ising}	M. Qin, Physica A \textbf{561}, 125176 (2021).
    
    \bibitem{QRG Khan} W. Joyia, S. Khan, K. Khan, Physica B: Condens. Mat. \textbf{601}, 412663 (2021). 	
    
    \bibitem{Q. C. spectrum} Y. C. Li, J. Zhang, and H.-Q. Lin, Phys. Rev. B \textbf{101}, 115142 (2020).
    
    \bibitem{Various Q. M.}	W.-Y. Sun, D. Wang, and L. Ye, Physica B: Condens. Mat. \textbf{524}, 27 (2017).  
    \bibitem{Triangular1} J.-Q. Cheng and J.-B. Xu, Phys. Rev. E \textbf{97}, 062134 (2018).  
      	
    
    \bibitem{Monte1} O. F. Syljuasen, Phys. Rev. A \textbf{68}, 060301(R) (2003).
    \bibitem{Monte2} O. F. Syljuasen, Phys. Lett. A \textbf{322}, 25 (2004).
    
    \bibitem{Monte3} T. Roscilde, P. Verrucchi, A. Fubini, S. Haas, and V. Tognetti, Phys. Rev. Lett. \textbf{94}, 147208 (2005).
    
    \bibitem{Monte4} L. X. Hayden, T. A. Kaplan, and S. D. Mahanti, Phys. Rev. Lett. \textbf{105}, 047203 (2010).
    \bibitem{Monte5} A. S. T. Pires, L. S. Lima, and M. E. Gouvea, J. Phys.: Condens. Matter \textbf{20}, 015208 (2008).
    \bibitem{Monte6} S. Gu, G. Tian, and H. Lin, Phys. Rev. A \textbf{71}, 052322 (2005).
 
    \bibitem{exactdiag} J. Song, S. Gu, and H. Lin, Phys. Rev. B \textbf{74}, 155119 (2006).
    
   	\bibitem{QRG Jafari} R. Jafari, A. Langari, Phys. Rev. B \textbf{76}, 014412 (2007).
    \bibitem{QRG XXZ}	M. Kargarian, R. Jafari, and A. Langari, Phys. Rev. A \textbf{77}, 032346 (2008).
    
    \bibitem{QRGG2} A. Langari, Phys. Rev. B \textbf{69}, 100402(R) (2004). 
    
      
    \bibitem{Q. E. 2D}	Y.-L. Xu, X.-M. Kong, Z.-Q. Liu, and C.-Y. Wang, Physica A \textbf{446}, 217 (2016).
    
    \bibitem{QRG XY} M. Usman, A. Ilyas, and K. Khan, Phys. Rev. A \textbf{92}, 032327 (2015). 
    \bibitem{Two-Dimensional XY Model} M. Usman and K. Khan, Eur. Phys. J.  D \textbf{74}, 1 (2020). 
  
	
	
	
		
	

    
    \bibitem{QUTIP1} J. R. Johansson, P. D. Nation, and F. Nori, Comput. Phys. Commun. \textbf{183}, 1760 (2012).
    
    \bibitem{QUTIP2}	J. R. Johansson, P. D. Nation, and F. Nori, Comput. Phys. Commun. \textbf{184}, 1234 (2013)
    
	
\end{thebibliography}
\end{document}